\documentclass[11pt]{article}
\usepackage[T1]{fontenc}
\usepackage{amssymb,amsmath,amsthm,amsfonts,dsfont,bm}
\usepackage{multirow}
\usepackage{graphicx}
\usepackage[hang,small,bf]{caption}
\usepackage{subcaption}
\usepackage{lscape}
\usepackage{color}
\usepackage{hyperref,url}
\usepackage[round,sort,comma,numbers]{natbib}

\usepackage{enumitem}
\usepackage[title]{appendix}
\usepackage{float}
\usepackage{mathtools}

\usepackage[english]{babel}
\usepackage[autostyle, english = american]{csquotes}
\MakeOuterQuote{"}

\usepackage{silence}
\definecolor{darkblue}{rgb}{0,0,0.55}
\newcommand{\dgreen}{\color[rgb]{0,0.6,0}}

\patchcmd{\appendices}{\quad}{: }{}{} 

\numberwithin{thm}{section}
\numberwithin{defn}{section}

\numberwithin{equation}{section}
\numberwithin{figure}{section}
\numberwithin{table}{section}

\newcommand{\ID}{\mathds{1}}

\hypersetup{colorlinks,allcolors=black}

\allowdisplaybreaks

\begin{document}   
\baselineskip 5mm

\thispagestyle{empty}

\begin{center}
{\LARGE Model Uncertainty and Selection of Risk Models for Left-Truncated and Right-Censored Loss Data
}

\vspace{12mm}

{\large\sc
Qian Zhao\footnote[1]
{
Qian Zhao, Ph.D., ASA, is
an Associate Professor in the Department of Mathematics, 
Robert Morris University, Moon Township, PA 15108, USA. 
~~ {\em e-mail\/}: ~{\tt zhao@rmu.edu}}}  

\vspace{1mm}
{\large\em Robert Morris University}
\vspace{10mm}

{\large\sc
Sahadeb Upretee\footnote[2]
{
Sahadeb Upretee, Ph.D., is
an Assistant Professor in the Department of Mathematics, 
Central Washington University, Ellensburg, WA 98926, USA. 
~~ {\em e-mail\/}: ~{\tt sahadeb.upretee@cwu.edu}}}  

\vspace{1mm}
{\large\em Central Washington University}
\vspace{10mm}

{\large\sc
Daoping Yu\footnote[3]
{
{\sc Corresponding Author:}
Daoping Yu, Ph.D., ASA, is
a Visiting Assistant Professor in the Department of Mathematical Sciences, 
University of Wisconsin-Milwaukee, Milwaukee, WI 53211, USA. 
~~ {\em e-mail\/}: ~{\tt dyu@uwm.edu}}}  

\vspace{1mm}
{\large\em University of Wisconsin-Milwaukee}
\vspace{12mm}

\end{center}

\vspace{12mm}

\noindent {\em\textbf{Abstract}.} 
 Insurance loss data are usually in the form of left-truncation and right-censoring due to deductibles and policy limits respectively. This paper investigates the model uncertainty and selection procedure when various parametric 
models are constructed to accommodate such left-truncated and right-censored data. The joint asymptotic properties of the estimators have been established using the Delta method along with Maximum Likelihood Estimation when the model is specified.  We conduct the simulation studies using Fisk, Lognormal, Lomax, Paralogistic, and Weibull distributions with various proportions of loss data below deductibles and above policy limits. A variety of graphic tools, hypothesis tests, and penalized likelihood criteria are employed to validate the models, and their performances on the model selection are evaluated
through the probability of each parent distribution being correctly selected. The effectiveness of each tool on model selection is also illustrated using {well-studied} data that represent Wisconsin
property losses in the
United States from 2007 to 2010.

\vspace{10mm}


\newpage
\section{Introduction}
In actuarial science, calculating financial and insurance risk measures plays an important role in the modeling of financial and insurance loss data. In many cases, loss data exhibits complex features such as left-truncation and right-censoring. These features pose challenges on model selection and the analysis of model uncertainty.\\

The selection of appropriate risk models for left-truncated and right-censored (LTRC) loss data is an essential task in actuarial practice. Selection of inaccurate models can lead to incorrect estimation of financial risk, resulting in sub-optimal decision-making. There is a need for a thorough analysis of model uncertainty and model selection for LTRC loss data. This article aims to address the model uncertainty and the corresponding effect on model selection. \\

In actuarial literature, \cite{Brazauskas2019} studied various aspects of LTRC insurance loss data such as probability density function (PDF), cumulative distribution function (CDF), quantile function (QF), building up a framework to derive asymptotic distributions of parametric and empirical estimators. 
Besides, \cite{poudyal_2023} designed some robust parametric estimation procedures for the insurance payment data affected by deductibles, policy limits, and coinsurance factors. These authors {clearly presented examples and showed}  how model uncertainty arises in the model fitting process of the actuarial loss data and how to deal with model mis-specification. \\

Left truncation and right censoring are both types of data incompleteness commonly encountered in survival analysis. \cite{Mitra2012} studied the left truncation and right censoring in lifetime data. Left truncation is the situation where the individuals in the study are only observed from a certain point in time forward, and those who have experienced the event of interest prior to this time point are excluded from the analysis. Right censoring refers to the situation where the event of interest has not occurred for some individuals by the end of the study period. In our study, left truncation means that losses below a certain deductible value have not been observed, and right censoring means that only frequencies are observed for losses above a certain policy limit value without further information about severity. This context of insurance loss data differs from the context of lifetime data for survival analysis. \\

In the model selection procedure, the selection criteria play a critical role in determining the best candidate model. The performance of typical penalized likelihood measures, Akaike Information Criterion (AIC) and Bayesian Information Criterion (BIC) have been reviewed (see \cite{Emura2022} for example) and discussed using various incomplete data sets with the data truncation and censoring facilitated according to their industrial applications. \cite{Balakrishnan2014} demonstrated that AIC is superior to the BIC when the true model comes from the Gamma, Lognormal and Weibull distributions. \cite{Mitra2021} suggested that the likelihood-based selection criterion is preferred to the distance-based selection criterion (such as Kolmogorov-Smirnov distance) within the Lehmann family distributions. We will also employ the Information Complexity (ICOMP) criterion, a measure that penalizes the interdependencies among parameter estimators of Maximum Likelihood Estimation, as a supplemental tool, in addition to AIC and BIC, to identify the structural differences among widely used medium and heavy tailed distributions. In this research, we compare all the above-mentioned selection indicators and consider the Anderson-Darling test, another goodness-of-fit test that puts more weight on
the tails, to evaluate the model performance on a wider range of loss models.\\

The remainder of the paper is organized as follows. In Section 2, we present key probabilistic and log-likelihood functions of the Fisk, Frechet, Lognormal, Lomax, Paralogistic, and Weibull distributions. In Section 3, we discuss Quantitile-Quantitile plot, Kolmogorov-Smirnov test and Anderson-Darling test for model validation. Further, AIC, BIC, and ICOMP criteria will be discussed to select the model in Section 4. The simulation study has been conducted in Section 5. A real data analysis has been conducted in Section 6. Furthermore, the main results of the paper are summarized in Section 7 {which concludes}.

\section{Theoretical Derivations for Candidate Distributions}
In this Section, we study six candidate loss distributions for medium and heavy tailed risks. They are Fisk, Frechet, Lognormal, Lomax, Paralogistic, and Weibull distributions. { Let $X^*$
represent the conditional random variable $X|d\leq X\leq u$.}

\subsection{Model Distribution with left truncation and right censoring}
\noindent
Candidate 1: Suppose random variable $X$ follows a Fisk distribution {(see \cite{Fisk1961} for example)} with scale parameter $\theta$ and shape parameter $\alpha$. The pdf and cdf for a ground-up Fisk loss distribution are: 
\begin{align*} \label{e4.2}
\text{PDF:} \quad f(x)&=\frac{\alpha \theta^{\alpha}x^{\alpha-1}}{(x^{\alpha}+\theta^{\alpha})^2}, \qquad x>0, {\alpha >0, \theta >0}\\
\text{CDF:}\quad  F(x) &=\frac{x^{\alpha}}{x^{\alpha}+\theta^{\alpha}},\qquad x>0 , {\alpha >0, \theta >0}
\end{align*}
This distribution is also called Loglogistic distribution. It resembles the Lognormal near zero when the shape parameter is greater than one and behaves more like Lomax in the tail.
 The cdf of $X^*$ representing the loss data with deductible ($d$) and policy limit ($u$) is:
\begin{align*}
F_{\ast}(x^*) 
& = 
\begin{cases}
0;
& \mbox{if}\quad x^* \leq d, \\[8pt]
\dfrac{F(x^*)-F(d)}{1-F(d)}=\dfrac{(x^*)^\alpha-d^\alpha}{(x^*)^\alpha+\theta^\alpha};
& \mbox{if} \quad d < x^* < u, \\[8pt] 
1 \quad & \mbox{if}\quad x^*\geq u.
\end{cases}
\end{align*}
Further, the qf of $X^*$ is:
\begin{align*}
F_{\ast}^{-1}(p)
& = 
\begin{cases}
\theta\left[\dfrac{d^{\alpha}+p\,\theta^{\alpha}}{(1-p)\,\theta^{\alpha}}\right]^{1/\alpha} 
& \mbox{for}\quad 0\leq p\leq p_{u}, \\[8pt]
u \quad & \mbox{for}\quad p_{u}\leq p\leq 1.
\end{cases}
\end{align*}
Note that $p_{u}=\dfrac{F(u)-F(d)}{1-F(d)}=\left(\dfrac{u}{\theta}\right)^{\alpha}\dfrac{d^{\alpha}+\theta^{\alpha}}{u^{\alpha}+\theta^{\alpha}}-\left(\dfrac{d}{\theta}\right)^{\alpha}$. Let $H(x^*)=(x^*){^{\alpha}}+\theta^{\alpha}$ , then
the log-likelihood function becomes
\begin{align*}
\log L
& = 
\sum_{i=1}^{n}\left[
\log f(x_{i}^*)\ID\{d<x_{i}^*<u\}+\log[1-F(u)]\ID(x_{i}^*=u)\right]-n\log\big(1-F(d)\big)\\
& = 
\sum_{i=1}^{n}\left[
\log \frac{\alpha\,\theta^{\alpha}\,x_{i}^{*(\alpha-1)}}{H^2(x_{i}^*)}\ID\{d<x_{i}^*<u\}+\log\frac{ \theta^{\alpha}}{H(u)}\ID\{x_{i}^*=u\}\right]
-n\log\frac{\theta^{\alpha}}{H(d)}\\
&=
\sum_{i=1}^{n}\bigg\{\big[\log \alpha+(\alpha-1)\log x^* -2\log H(x_{i}^*)\big]\ID\{d<x_{i}^*<u\}-\log H(u)\ID\{x_{i}^*=u\}\bigg\}
+n\log H(d).
\end{align*}

\vspace{12pt}

\noindent
Candidate 2: The Frechet, known as inverse Weibull distribution, is a special case of the generalized extreme value distribution with much fatter tails on the right. It has ground-up pdf and cdf as follows
\begin{align*}
\text{PDF:} \quad f(x)&=\frac{\alpha (\theta/x)^{\alpha}e^{-(\theta/x)^{\alpha}}}{x}, \qquad x>0,  \alpha>0, \theta>0,\\
\text{CDF:}\quad  F(x)&=e^{-(\theta/x)^{\alpha}}, \qquad x>0, \theta>0.
\end{align*}
Then cdf of $X^*$ representing the loss data with deductible (d) and policy limit (u) is:
\begin{align*}
F_{\ast}(x^*) 
& = 
\begin{cases}
0;
& \mbox{if}\quad x^* \leq d, \\[8pt]
\dfrac{F(x^*)-F(d)}{1-F(d)}=\dfrac{e^{-(\theta/x^*)^{\alpha}}-e^{-(\theta/d)^{\alpha}}}{1-e^{-(\theta/d)^{\alpha}}},
& \mbox{if} \quad d < x^* < u, \\[8pt] 
1, \quad & \mbox{if}\quad x^*\geq u.
\end{cases}
\end{align*}
Further, the qf of $X^*$ is:
\begin{align*}
F_{\ast}^{-1}(p)
& = 
\begin{cases}
\theta\left[-\log \left(p+(1-p)e^{-(\theta/d)^{\alpha}}\right)\right]^{-1/\alpha},
& \mbox{for}\quad 0\leq p\leq p_{u}, \\[8pt]
u, \quad & \mbox{for}\quad p_{u}\leq p\leq 1.
\end{cases}
\end{align*}
 Note that $p_{u}=\dfrac{F(u)-F(d)}{1-F(d)}=\dfrac{e^{-(\theta/u)^{\alpha}}-e^{-(\theta/d)^{\alpha}}}{1-e^{-(\theta/d)^{\alpha}}}$. Let $S(x)=1-F(x)=1-e^{-(\theta/x)^{\alpha}}$, the log-likelihood function becomes
\begin{align*}
\log L
& = 
\sum_{i=1}^{n}\left[
\log f(x_{i}^*){\ID\{d < x^* < u\}}+\log[1-F(u)]{\ID\{ x^* = u\}}\right]-n\log\big[1-F(d)\big]\\
& = 
\sum_{i=1}^{n}\left[
\log [\alpha \theta^{\alpha}x^{*(-\alpha-1)}e^{-(\theta/x^*)^{\theta}}]{\ID\{d < x^* < u\}}+\log S(u){\ID\{ x^* = u\}}\right]
-n\log S(d)\\
&=
\sum_{i=1}^{n}\bigg \{\big[\log \alpha+\alpha\log \theta-(\alpha+1)\log x^*-(\theta/x^{*\alpha}\big]{\ID\{d < x^* < u\}}+\log S(u){\ID\{ x^* = u\}}\bigg \}-n\log S(d).
\end{align*}

\vspace{12pt}
\noindent
Candidate 3:
Suppose random variable $X$ is distributed according to a Lognormal distribution with log-location parameter $-\infty < \mu < \infty$, and log-scale parameter 
$\sigma > 0$. We will denote this fact as 
$X \sim \mathcal{LN} \left( \mu, \sigma \right)$. Its pdf and cdf are:
\begin{align*}
\text{PDF:}\quad f(x)  & ~=~ (\sigma (x))^{-1} \,
\varphi \left( \frac{\log (x)-\mu}{\sigma} \right), 
\qquad x > 0,\\
\text{CDF:}\quad F(x) &  ~=~ 
\Phi \left( \frac{\text{log}(x) -\mu}{\sigma}\right), 
\qquad x > 0.
\end{align*}
Here $\Phi$, $\varphi$, $\Phi^{-1}$ denote the cdf, pdf, qf of 
the standard normal distribution, respectively. Next, let us introduce the following abbreviations:
\[
c_d ~:=~ \frac{\log(d) - \mu}{\sigma}, 
\qquad
c_x ~:=~ \frac{\log(x) - \mu}{\sigma},
\qquad
c_u ~:=~ \frac{\log(u) - \mu}{\sigma}.
\]
 Then cdf of $X^*$ representing the loss data with deductible and policy limit is
\begin{eqnarray*}
F_*(x^*|d;u) = 
\left\{
\begin{array}{ll}
 0, & \quad x^* \leq d, 
\\[2ex]
{\displaystyle
\frac{\Phi \left( c_{x^*} \right) - \Phi \left( c_d \right)}
{1 - \Phi \left( c_d \right)}, 
} & \quad d < x^* < u,
\\[2ex]
 1, & \quad x^* \geq u.
\end{array}
\right.
\label{LN2}
\end{eqnarray*}
Further, the qf of $X^*$ is:
\begin{eqnarray*}
F_*^{-1} (p|d;u) = 
\left\{
\begin{array}{ll}
\exp 
\left\{ 
\mu + \sigma \Phi^{-1} 
\Big( p + (1-p) 
\Phi \left( c_d \right) 
\Big) 
\right\},
& \quad  0 \leq p < p_u, 
\\[0.5ex]
 u, & \quad  p_u \leq p \leq 1.
\end{array}
\right.
\label{LN3}
\end{eqnarray*}
Note that
$p_u = [\Phi(c_u) - \Phi(c_d)] / [1 - \Phi(c_d)]$. Finally, the log-likelihood function is
\begin{align*}
\log {\cal{L}}
& =
 \sum_{i=1}^n 
\big[ 
-\log(\sigma) - \log(x_{i}^*) + \log(\varphi(c_{x_{i}^*}))
\big] \ID\{ d < x_{i}^* < u \big\} + \log \left[ 1 - \Phi(c_u) \right] 
\sum_{i=1}^n \ID\{ x_{i}^* = u \big\}
\\
 &\quad - n \log \left[ 1 - \Phi(c_d) \right].
\end{align*}
\vspace{12pt}

\noindent
Candidate 4: Suppose random variable $X$ is distributed according to Lomax distribution {(see \cite{Lomax1954} for example)} (also known as Pareto Type II, {see \cite{Arnold2015} for example}). The pdf and cdf for a ground-up Lomax loss distribution are:

\begin{align*}
\text{PDF:}& \quad f(x)=\frac{\alpha \theta^\alpha}{(x+\theta)^{\alpha+1}}, \quad x>0,\\
\text{CDF:}& \quad F(x)=1-\left(\frac{\theta}{x+\theta}\right)^\alpha, \quad x>0. 
\end{align*}
Then cdf of $X^*$ representing the loss data with deductible (d) and policy (u) limit is:
\begin{align*}
F_{\ast}(x^*|d;u) 
& = 
\begin{cases}
0; 
& \mbox{if}\quad x^* \leq d, \\[12pt]
{\displaystyle 1-\left(\frac{d+\theta}{x^*+\theta}\right)^{\alpha}}; 
& \mbox{if} \quad d < x^* < u, \\[8pt] 
1, \quad & x^*\geq u .
\end{cases}
\end{align*}
Further, the qf of $X^*$ is:
\begin{align*}
F_{\ast}^{-1}(p|d;u)
& = 
\begin{cases}
(1-p)^{-1/\alpha}(d+\theta)-\theta ,
& \mbox{if}\quad 0\leq p < p_{u}, \\[8pt]
u, \quad & \mbox{if}\quad p_{u}\leq p\leq 1.
\end{cases}
\end{align*}
Note that $p_{u}=1-\left(\frac{d+\theta}{u+\theta}\right)^{\alpha}$.
The log-likelihood function becomes
\begin{align*}
\log {\cal{L}}
& = \sum_{i=1}^{n}\left[\log\alpha -(\alpha+1)\log(x_{i}^*+\theta)\right]\ID\{d<x_{i}^*<u\} -\alpha \log(u+\theta)\ID\{x_{i}^*=u\}+ \alpha \log(d+\theta).
\end{align*}
\vspace{12pt}

\noindent
Candidate 5: Suppose random variable $X$ is distributed according to Paralogistic distribution. The pdf and cdf for a ground-up Paralogistic loss distribution are:

\begin{align*}
\text{PDF:} &\quad
f(x)=\frac{\alpha^2 x^{\alpha-1}\theta^{\alpha^2}}{\left(\theta^\alpha+x^\alpha\right)^{\alpha+1}}, \qquad x>0, \\
\text{CDF:} &\quad
F(x)=1-\left(\frac{\theta^\alpha}{\theta^\alpha+x^\alpha}\right)^\alpha, \qquad x>0.
\end{align*}
\noindent
Then cdf of $X^*$ representing the loss data with deductible (d) and policy (u) limit is:
\begin{align*}
F_{\ast}(x^*|d;u) 
& = 
\begin{cases}
0 ,
& \mbox{if}\quad x^*\leq d, \\[8pt]
\dfrac{F(x^*)-F(d)}{1-F(d)}=1-\left(\dfrac{\theta^\alpha+d^\alpha}{\theta^\alpha+x^{*\alpha}}\right)^\alpha, 
& \mbox{if} \quad d < x^* < u, \\[8pt] 
1, \quad & \mbox{if}\quad x^* \geq u.
\end{cases}
\end{align*}
Further, the qf of $X^*$ is:
\begin{align*}
F_{\ast}^{-1}(p|d;u)
& = 
\begin{cases}
F^{-1}\big(p+(1-p)F(d)\big)=\left(\dfrac{\theta^\alpha+d^\alpha-\theta^\alpha(1-p)^{1/\alpha}}{(1-p)^{1/\alpha}}\right)^{1/\alpha}, 
& \mbox{for}\quad 0\leq p < p_{u}, \\[8pt]
u, \quad & \mbox{for}\quad p_{u}\leq p\leq 1.
\end{cases}
\end{align*}
Note that $p_{u}=\dfrac{F(u)-F(d)}{1-F(d)}=1-\left(\dfrac{\theta^\alpha+d^\alpha}{\theta^\alpha+u^\alpha}\right)^\alpha$. Then the log-likelihood function becomes
\begin{align*}
\log {\cal{L}}
& = 
\sum_{i=1}^{n}\bigg\{
\log f(x_i^*)\ID\{(d<x_{i}^*<u)\}+\Big[\alpha^2\log\theta-\alpha\log\left(\theta^\alpha+u^\alpha\right)\Big]\ID\{x_{i}^*=u\}\bigg\}-n\Big[\alpha^2\log\theta-\alpha\log\left(\theta^\alpha+d^\alpha\right)\Big]\\
&=
\sum_{i=1}^{n}\bigg\{
\Big[2\log\alpha+(\alpha-1)\log x_{i}^*-(\alpha+1)\log\left(\theta^\alpha+x_{i}^{*\alpha}\right)\Big]\ID\{(d<x_{i}^*<u)\}-\alpha\log\left(\theta^\alpha+u^\alpha\right)\ID\{x_{i}^*=u\}\bigg\}\\
&\quad+n\,\alpha\log\left(\theta^\alpha+d^\alpha\right).
\end{align*}
\vspace{12pt}

\noindent
Candidate 6: Suppose random variable $X$ follows a Weibull distribution. The pdf and cdf for a ground-up Weibull loss distribution are:

\begin{align*}
\text{PDF:} &\quad
f(x)=\frac{\alpha}{x} \left(\frac{x}{\theta}\right)^{\alpha}e^ {- \left( x/ \theta \right)^{\alpha}}, \qquad x>0, \\
\text{CDF:} &\quad
F(x)=1-e^ {- \left( x/ \theta \right)^{\alpha}}, \qquad x>0.
\end{align*}
Then cdf of $X^*$ representing the loss data with deductible (d) and policy (u) limit is:
\begin{align*}
F_{\ast}(x^*|d;u) 
& = 
\begin{cases}
0 ,
& \mbox{if}\quad x^* \leq d, \\[8pt]
\dfrac{F(x^*)-F(d)}{1-F(d)}=1-e^{\left(d/\theta\right)^{\alpha}-\left(x^*/\theta\right)^{\alpha}} ,
& \mbox{if} \quad d < x^* < u, \\[8pt] 
1, \quad & \mbox{if}\quad x^* \geq u.
\end{cases}
\end{align*}
\noindent
Further, the qf of $X^*$ is:
\begin{align*}
F_{\ast}^{-1}(p|d;u)
& = 
\begin{cases}\theta\left(-\log(1-p)+\left(\frac{d}{\theta}\right)^{\alpha}\right)^{1/\alpha}, 
& \mbox{for}\quad 0\leq p < p_{u}, \\[8pt]
u, \quad & \mbox{for}\quad p_{u}\leq p\leq 1.
\end{cases}
\end{align*}
Note that $p_{u}=\dfrac{F(u)-F(d)}{1-F(d)}=1-e^{\left(d/\theta\right)^{\alpha}-\left(u/\theta\right)^{\alpha}}$
\noindent
. Then the log-likelihood function is
\begin{align*}
\log {\cal{L}}
& = 
\sum_{i=1}^{n}\bigg\{
\log f(x_{i}^*)\ID\{d<x_{i}^*<u)\}-\left(\frac{u}{\theta}\right)^{\alpha}\ID\{x_{i}^*=u\}\bigg\}+n\left(\frac{d}{\theta}\right)^{\alpha}\\
&=
\sum_{i=1}^{n}\bigg\{
\left[\log\alpha-\log x_{i}^*+\alpha\log x_{i}^*-\alpha\log\theta-\left(\frac{x_{i}^*}{\theta}\right)^\alpha\right]\ID\{d<x_{i}^*<u)\}-\left(\frac{u}{\theta}\right)^{\alpha}\ID\{x_{i}^*=u\}\bigg\}+n\left(\frac{d}{\theta}\right)^{\alpha}.
\end{align*}

\subsection{MLE Approach}

Parametric methods use the observed data $x^*_1, \ldots, x^*_n$ and fully 
recognize {\dgreen their} distributional properties. The Maximum Likelihood Estimation (MLE) approach is one of the most 
common estimation techniques.  Parameter estimates are found by maximizing the following log-likelihood 
function:

\vspace{-5mm}

\[
\log {\cal{L}} \big( \boldsymbol{\theta} \, \big| \, x^*_1, \ldots, x^*_n \big) 
 ~=~ 
\log \left[ \prod_{i=1}^n f_* (x^*_i \, | \, d; u) \right]  
 ~=~ \log \left[ 
\prod_{i=1}^n 
\left[ \frac{f(x^*_i)}{1-F(d)} \right]^{{\dgreen \ID }\{ d < x^*_i < u \}}
\left[ 
\frac{1 - F(u^-)}{1-F(d)} \right]^{{\dgreen \ID} \{ x^*_i = u \}}
\right]
\]

\vspace{-5mm}

\begin{equation*}
 =~ \sum_{i=1}^n \log \big[ f(x^*_i) \big] {\dgreen \ID} \{ d < x^*_i < u \} 
~-~ n \log \big[ 1 - F(d) \big]
 ~+~
\log \big[ 1 - F(u^-) \big] \sum_{i=1}^n {\dgreen \ID} \{ x^*_i = u \},
\label{logL}
\end{equation*}
where ${\dgreen \ID} \{\cdot\}$ denotes the indicator function {and $u^{-}$ represents a point closing to $u$ from the left}.\\

Once parameter MLEs, $\widehat{\theta}_1, \ldots, \widehat{\theta}_k$, are 
available, the $p$th quantile estimate is found by plugging those MLE values 
into the parametric expression of $F^{-1}(p) = h(\theta_1, \ldots, \theta_k)$,  where $h(\theta_1, \ldots, \theta_k)$ is a function of parameters. 

\subsection{Fisher Information}
Let $\widehat{\mbox{\boldmath $\theta$}} = 
\left( \widehat{\theta}_1, \ldots, \widehat{\theta}_k \right)$ 
denote the MLE of parameter 
$\mbox{\boldmath $\theta$} = (\theta_1, \ldots, \theta_k)$. 
Then, the Fisher information matrix ${\mbox{\boldmath $I$}}$, with the entries given by
\begin{equation*}
I_{ij} = \mbox{\bf E} 
\left[
\frac{\partial \log g(X)}{\partial \theta_i} \cdot 
\frac{\partial \log g(X)}{\partial \theta_j} 
\right],
\end{equation*}
where $\mbox{\bf E}[\cdot]$ denotes the expectation operator, and $g$ is the true underlying pdf. \\

the observed Fisher information represents the amount of information that the observed data provides about the parameter $\boldmath{ \theta}$.
Intuitively, a high Fisher information indicates that the data {contain} valuable information about the parameter, and it implies that the estimate of the parameter based on the data is likely to be more precise. On the other hand, a low Fisher information suggests that the data provides little information about the parameter, leading to higher uncertainty in the estimate. The variance-covariance matrix is related to the inverse of the Fisher information matrix.

\section{Model Validation}
 A variety of graphic tools, hypothesis tests, and penalized likelihood criteria are employed to select the model and validate the performance.
\subsection{Quantile-Quantile Plot}
A Quantile-Quantile (QQ) plot is a graphical representation used to compare two probability distributions. To construct a QQ plot, we plot the sample distribution on the $Y-$axis and the theoretical distribution on the $X-$axis. When the points in the QQ plot align closely to a diagonal line at 45 degrees, it indicates a high level of agreement between the sample distribution and the theoretical distribution.\\

In order to avoid visual distortions due to large spacings between the most extreme observations, both axes can be measured on the logarithmic scale. The QQ plot is generated using the following formula.
\[
\left( 
\log \left( 
\widehat{F}^{-1} \big[ u_{i} + \widehat{F}(d) (1-u_{i}) \big] 
\right), \; 
\log \left( x^*_{(i)} \right) 
\right),  ~~~ i = 1, \ldots, n,
\]
where $\widehat{F}(d)$ is the estimated parametric cdf evaluated 
at $d$, $\widehat{F}^{-1}$ is the estimated parametric qf,
$x^*_{(1)} < \cdots < x^*_{(n)}$ denote the ordered claim severities, 
$u_{i} = (i-0.5)/n$ is the quantile level.

\subsection{Kolmogorov-Smirnov Test and Anderson-Darling Test}
Besides the graphical visual check, multiple quantitative hypothesis tests can help compare the model performance and specify an appropriate fitted parametric distribution. Here Kolmogorov-Smirnov (KS) Test and Anderson-Darling (AD) Test are considered in the following model validation procedure.\\

For a KS test, the test statistic is
$$KS = \max_{d\leq x\leq u}|F_{n}(x)-{\widehat{F}_*(x)}| ,$$
where $d$ is the left truncation point ($d = 0$ if there is no truncation) and $u$ is the right
censoring point ($u = \infty$ if there is no censoring). \\

AD Test is similar to the KS test but uses a different measure of the
difference between the two distribution functions. The test statistic is
$$AD = n\int_{d}^{u}\frac{[F_{n}(x)-{\widehat{F}_*(x)}]^2}{{\widehat{F}_*(x)}[1-{\widehat{F}_*(x)}]}f^{\ast}(x)\,dx.$$
The AD goodness-of-fit statistic measures 
the cumulative weighted quadratic
distance (with more weight on the tails) between the empirical 
cdf and the parametrically estimated cdf. The computational formula for LTRC data is given by
\begin{align*}
\mbox{AD}_n& ~=~ -n \widehat{F}_*(u^-) + n \sum_{i=1}^k \big( i/n \big)^2
\log \left(
\frac{\widehat{F}_*(X_{(i+1)})}{\widehat{F}_*(X_{(i)})}
\right)
-
n \sum_{i=0}^{k-1}
\big( 1 - i/n \big)^2
\log \left(
\frac{1-\widehat{F}_*(X_{(i+1)})}{1-\widehat{F}_*(X_{(i)})}
\right)\\
&
-
\dfrac{(n-k)^2}{n} 
\log \left(
\dfrac{1-\widehat{F}_*(u^-)}{1-\widehat{F}_*(X_{(k)})}
\right),
\end{align*}
where the unique noncensored data points are $d=x_{0}<x_{1}\cdots<x_{k}<x_{k+1}=u$ and $\widehat{F}_*(u^-)=\big(\widehat{F}(u)-\widehat{F}(d)\big)/ \big(1-\widehat{F}(d)\big)$ (See more details in \cite{klugman2019loss}, section 15.4). \\

\section{Model Selection Criteria and Validation methods}
\subsection{Selection Criteria: AIC, BIC, and ICOMP}
Akaike Information Criteria $(AIC)$ is a statistical measure that can be used to assess and compare the goodness-of-fit and complexity of different statistical models. $AIC$ is a useful tool to compare different models in the context of model selection and is relevant when dealing with multiple candidate models. Akaike (1973) has provided formula as 
\begin{equation*}
   AIC = -2\log f \big( \boldsymbol x_n \, \big| \, \hat{\boldsymbol\theta}_p \big) + 2p, 
\end{equation*}
where $p$ is the number of parameters. Bayesian Information Criteria $(BIC)$ is another statistical measure which is used to assess quality of the goodness-of-fit and selection of models. Schwarz (1978) has provided formula to compute $BIC$ as
\begin{equation*}
     BIC=-2\log f \big( \boldsymbol x_n \, \big| \, \hat{\boldsymbol\theta}_p \big) + p\cdot\log(n).
\end{equation*}
Similar to AIC, the model with the lowest BIC value is considered the most appropriate choice. It balances the trade-off between the model accurately capturing the data and employing a concise number of model parameters. Moreover, similar to $AIC$ and $BIC$, \cite{Bozdogan1987, Bozdogan1990} has introduced another metric Information Complexity (ICOMP) and its corresponding value is given by  
\begin{equation*}
  ICOMP=  -2\log f \big( \boldsymbol x_n \, \big| \, \hat{\boldsymbol\theta}_p \big)  
+ 2C\left(\boldsymbol{\Sigma}(\hat{\boldsymbol\theta}_p) \right)
\end{equation*}
with $$C\left(\boldsymbol{\Sigma}(\hat{\boldsymbol\theta}_p) \right) 
= \frac{s}{2}\log\left(\frac{\mbox{tr}\left(\boldsymbol{\Sigma}(\hat{\boldsymbol\theta}_p) \right)}{s} \right) 
- \frac{1}{2}\log\left(\mbox{det}\left(\boldsymbol{\Sigma}(\hat{\boldsymbol\theta}_p) \right) \right) \nonumber,$$ with $s$, tr, and det denoting the rank, trace, and determinant of $\boldsymbol{\Sigma}$ 
(the variance-covariance matrix), respectively.

\subsection{Validation Methods}
As we know, the smaller the value of a likelihood-based measure is, the better the performance of a parametric loss model. However,
how much difference in a likelihood-based measure really matters and separates the models? Suppose we set up a null hypothesis $H_{0}:$ Model $i$ is the parent candidate that fits a data set and use BIC to select the candidate. \cite{Fabozzi2014}, p. 403 defined  $\Delta_{i}=BIC_{i}-BIC_{\text{min}}$, the difference of BIC between the model $i$ and the model with the smallest BIC, and then used the range of $\Delta_{i}$ (see Table \ref{tab:BIC}) to indicate the strength of evidence to take against the model $i$ as the parent distribution.
\begin{table}[htb]
    \centering
    \begin{tabular}{c|c}
    \hline
      Range of $\Delta_{i}$  & Evidence against a candidate\\
      \hline
       $[0,2]$  &  little\\
       $(2,6]$ & positive\\
        $(6,10]$ & strong\\
        $(10,\infty)$ & very strong\\
         \hline
    \end{tabular}
\caption{Strength of evidence to take against the model $i$}\label{tab:BIC}
\end{table}

From the definition, the value of BIC highly depends on the sample size $n$, the larger the sample size, the higher the scale of a BIC could be. Therefore, for the same difference $\Delta_{i}=10$, it is relatively easier to separate fitted models for a data set with sample $n=1,000$ rather than a data set with $n=10,000$. To avoid such an impact of sample size on model selection and validation, we report two types of probabilities instead of the pure BIC values (or the differences) in the following simulation study.
\begin{itemize}
    \item Type 1 -- the probability of each parent distribution being correctly selected. Inspired by \cite{Mitra2021}, we can fit all the candidate models including the parent one to an LTRC data, and then choose the model with the lowest BIC number as the best ﬁt. We repeat this process $N$ times and calculate the proportion of times each candidate is chosen as the best, resulting in the probability for each of the candidate models to be selected. The same process is followed for other likelihood-based criteria AIC and ICOMP and the distance-based measure KS statistic and AD statistic. 
    \item Type 2 -- the posterior model probabilities. Suppose we only have $k$ models at hand and assume that at the beginning, each model has the same chance ($1/k$) being the underlying distribution. Then the difference in likelihood can update the possibility of a  model being selected and provides an insight about the best candidate model.  Defined in \cite{Raftery1995} and \cite{Kenneth2004}, the posterior probability of model $i$ is
$$p_{i}=\frac{\exp(-\frac{1}{2}\Delta_{i})}{\sum_{j=1}^{k} \exp(-\frac{1}{2}\Delta_{j})}.$$
Here, the constant $1/2$ in exponent will balance the multiplier $2$ in the BIC definition and the ratio structure will reduce the impact of sample size on BIC comparison.  The posterior probability of AIC and ICOMP can be calculated using the same formula, and we just need to change the term $\Delta$ as the difference for those two criteria. 
\end{itemize}

\section{Simulation Study}

In this section, we conduct an analysis of model uncertainty due to both left truncation and right censoring structure and investigate the effects of various combinations of proportions on the model selection for the above-mentioned loss models. We fix the deductible $d$ and the policy limit $u$ for all models and tune the parameter values of each model so that $F(d)$ and $F(u)$ are both pre-specified constants. The parameter values can be determined using the percentile matching at two percentile levels, namely $F(d)$ and $F(u)$.
The data are then generated according to those parameter values of an underlying distribution and the MLE estimations are conducted to fit the data to each of the candidate models. Multiple model selection rules are implemented to specify the parent distribution of data and their performance are compared in terms of selection accuracy and likelihood probabilities ultimately.\\

The simulation study is designed with the following setups:
\begin{itemize}
    \item[(i)] Five selected loss models: Fisk, Lognormal, Lomax, Paralogistic, Weibull
    \item[(ii)] Sample size: $n=1,000$ and $100,000$
    \item[(iii)]  Truncation and censoring thresholds: $d=500, u=10,000$
   \item[(iii)]  Truncation and {non-}censoring probabilities $(F(d), F(u))$: (0.10, 0.85) and (0.50, 0.80) 
   \item[(iv)] Model selection indicators: KS statistic, AD statistic, AIC, BIC, ICOMP
\end{itemize}
{\emph {NOTE}}: Frechet is not included in the simulation study due to the instability of parameter estimation.\\

{The choice of sample size, truncation and censoring thresholds, truncation probability and censoring probability depends on practical considerations and research objectives. A sample size of 1,000 can be considered moderate for many statistical studies. It is often chosen due to constraints on resources. It can provide a reasonable level of statistical power to detect meaningful effects. A sample size of 100,000 is quite large and may be chosen for studies where researchers need to achieve a very high level of statistical power, detect small effect sizes, and adequately capture rare events. A deductible of 500 and a policy limit of 10,000 are very common in insurance industry, which provides a practical foundation for our choice of truncation threshold being 500 and censoring threshold being 10,000. A truncation probability of 10\% is kind of low in insurance, which means not too much information loss due to truncation but is useful for comparison with another scenario of more information loss due to a higher truncation probability. A truncation probability of 50\% is high in theory but quite ordinary in insurance practice due to the fact of many loss events with magnitudes below deductibles. Both a censoring probability of 15\% and that of 20\% imply that a substantial portion of the subjects experience censoring during the study. This could be chosen based on prior knowledge or to simulate realistic scenarios. The choice of these parameters depends on the balance between theoretical rigor, practical considerations, and the objectives of the study. \\ }

\begin{table}[hbt!]
{\footnotesize  
\caption{$100$ simulations of Fisk data ($\alpha=1.31$, $\theta=2,667$); Logn data ($\mu=7.87$, $\theta=1.29$); Lomax data ($\alpha=8.69$, $\theta=41,007$); Paralog data ($\alpha=1.24$, $\theta=3,533$); Weibull data ($\alpha=0.96$, $\theta=5,150$); $d=500$, $u=10,000$, $n=1,000$. $F(d)=0.10$, $F(u)=0.85$.}
\label{tab:1}
\begin{tabular}{c|c|c|c|c|c|c}
\hline
\multicolumn{2}{c|}{} &
\multicolumn{1}{|c|}{Fisk} &
\multicolumn{1}{|c|}{Logn} &
\multicolumn{1}{|c|}{Lomax} &
\multicolumn{1}{|c|}{Paralog} &
\multicolumn{1}{|c}{Weibull} \\
\hline\hline
\multirow{6}{4em}{Fisk Data} & KS  &  $0.54^{\ast}$ & 0.15 & 0.10 & 0.21 & 0.00\\ 
 & AD  &  0.56 & 0.15 & 0.06 & 0.23 & 0.00\\ 
& AIC  &  0.72 $(0.38, 0.36)^{\ast\ast}$ & 0.19 (0.21, 0.20) & 0.03 (0.12, 0.10) & 0.06 (0.29, 0.30) & 0.00 (0.01, 0.00)\\ 
& BIC   &  0.72 (0.37, 0.36) & 0.18 (0.21, 0.20) & 0.05 (0.12, 0.10) & 0.06 (0.29, 0.30) & 0.00 (0.01, 0.00)\\ 
& ICOMP &  0.00 (0.00, 0.00) & 1.00 (1.00, 1.00)& 0.00 (0.00, 0.00) & 0.00 (0.00, 0.00)& 0.00 (0.00, 0.00)\\ 
\hline\hline
\multirow{6}{4em}{Logn Data} & KS & 0.26  & 0.16 & 0.27 & 0.18 & 0.13  \\ 
& AD  & 0.27 &  0.30 & 0.17 & 0.15 & 0.11\\ 
& AIC & 0.33 (0.20, 0.16) & 0.43 (0.26, 0.27) & 0.12 (0.23, 0.23) & 0.03 (0.21, 0.23) & 0.09 (0.11, 0.02) \\ 
& BIC & 0.33 (0.19, 0.16) & 0.44 (0.26, 0.26) & 0.10 (0.23, 0.23) & 0.02 (0.20, 0.23) & 0.11 (0.11, 0.02) \\ 
& ICOMP &  0.00 (0.00, 0.00) & 1.00 (1.00, 1.00) & 0.00 (0.00, 0.00) & 0.00 (0.00, 0.00) & 0.00 (0.00, 0.00) \\
\hline\hline
\multirow{6}{4em}{Lomax Data} & KS & 0.00  & 0.02 & 0.42 & 0.14 & 0.42  \\ 
& AD  & 0.00  & 0.02 & 0.39 & 0.14 & 0.45  \\ 
& AIC & 0.00 (0.02, 0.00) & 0.06 (0.05, 0.01) & 0.73 (0.45, 0.48) & 0.06 (0.09, 0.02) & 0.15 (0.40, 0.48) \\ 
& BIC & 0.00 (0.02, 0.00) & 0.06 (0.05, 0.01) & 0.68 (0.45, 0.48) & 0.05 (0.09, 0.02) & 0.21 (0.40, 0.47) \\ 
& ICOMP & 0.00 (0.00, 0.00) & 0.93 (0.86, 0.98) & 0.03 (0.04, 0.00) & 0.04 (0.00, 0.01) & 0.05 (0.10, 0.01) \\ 
\hline\hline
\multirow{6}{4em}{Paralog Data} & KS  & 0.39  & 0.22 & 0.17 & 0.21 & 0.01  \\ 
& AD  & 0.37  & 0.29 & 0.12 & 0.22 & 0.00 \\ 
& AIC & 0.53 (0.29, 0.29) & 0.27 (0.26, 0.26) & 0.11 (0.16, 0.12) & 0.09 (0.28, 0.28) & 0.00 (0.02, 0.00) \\ 
& BIC & 0.51 (0.29, 0.27) & 0.30 (0.26, 0.27) & 0.07 (0.15, 0.12) & 0.12 (0.28, 0.29) & 0.00 (0.02, 0.00) \\ 
& ICOMP & 0.00 (0.00, 0.00) & 1.00 (1.00, 1.00) & 0.00 (0.00, 0.00) & 0.00 (0.00, 0.00) & 0.00 (0.00, 0.00) \\ 
\hline\hline
\multirow{6}{4em}{Weibull Data} & KS  & 0.00  & 0.01 & 0.32 & 0.08 & 0.59  \\ 
& AD  & 0.00  & 0.00 & 0.42 & 0.08 & 0.50 \\ 
& AIC & 0.00 (0.01, 0.00) & 0.00 (0.02, 0.00) & 0.75 (0.46, 0.50) & 0.05 (0.06, 0.01) & 0.20 (0.46, 0.50) \\ 
& BIC & 0.00 (0.01, 0.00) & 0.00 (0.02, 0.00) & 0.72 (0.45, 0.49) & 0.05 (0.06, 0.01) & 0.23 (0.46, 0.50) \\ 
& ICOMP & 0.00 (0.00, 0.00) & 0.45 (0.61, 0.82) & 0.21 (0.21, 0.00) & 0.00 (0.00, 0.00) & 0.14 (0.18, 0.02) \\ 
\hline
\end{tabular}
\\[8pt]
NOTE: $\ast$ Probability of selecting a model in various parent data distributions. $\ast\ast$ The likelihood-based approach has additional posterior probability in parenthesis $(a,b)$, where $a$ represents the mean, and $b$ is the median.}
\end{table} 

 Tables \ref{tab:1} - \ref{tab:4} demonstrate how often LTRC data can correctly identify its underlying distribution with various combinations of left-truncation and right-censoring proportions and how the sample size affects correct identification. The selecting probabilities of a model within all indicator measurements (KS, AD, AIC, BIC, and ICOMP) are recorded as the first value in each cell and the likelihood-based approach also reports posterior probability in additional parenthesis, where the former number is the mean of 100 simulations while the latter one is the median. The computation is run through {\emph{Matlab}}.\\

\begin{table}[hbt!]
{\footnotesize \caption{$100$ simulations of Fisk data ($\alpha=1.31$, $\theta=2,667$); Logn data ($\mu=7.87$, $\theta=1.29$); Lomax data ($\alpha=8.69$, $\theta=41,007$); Paralog data ($\alpha=1.24$, $\theta=3,533$); Weibull data ($\alpha=0.96$, $\theta=5,150$); $d=500$, $u=10,000$, $n=100,000$. $F(d)=0.10$, $F(u)=0.85$.}
\label{tab:2}
\begin{tabular}{c|c|c|c|c|c|c}
\hline
\multicolumn{2}{c|}{} &
\multicolumn{1}{|c|}{Fisk} &
\multicolumn{1}{|c|}{Logn} &
\multicolumn{1}{|c|}{Lomax} &
\multicolumn{1}{|c|}{Paralog} &
\multicolumn{1}{|c}{Weibull} \\
\hline\hline
\multirow{6}{4em}{Fisk Data} & KS  &  0.96 & 0.00 & 0.00 & 0.04 & 0.00\\ 
& AD  &  0.96 & 0.00 & 0.00 & 0.04 & 0.00\\ 
& AIC  &  0.96 (0.96, 1.00) & 0.00 (0.00, 0.00) & 0.00 (0.00, 0.00) & 0.04 (0.04, 0.04) & 0.00 (0.00, 0.00)\\ 
& BIC   &  0.96 (0.96, 1.00) & 0.00 (0.00, 0.00) & 0.00 (0.00, 0.00) & 0.04 (0.04, 0.04) & 0.00 (0.00, 0.00)\\ 
& ICOMP &  0.97 (0.96, 1.00) & 0.00 (0.00, 0.00)& 0.00 (0.00, 0.00) & 0.03 (0.04, 0.00)& 0.00 (0.00, 0.00)\\ 
\hline\hline
\multirow{6}{4em}{Logn Data} & KS & 0.00  & 0.94 & 0.05 & 0.01 & 0.00  \\ 
& AD  & 0.00  &  0.99 & 0.01 & 0.00& 0.00\\ 
& AIC & 0.00 (0.00, 0.00) & 0.99 (0.98, 1.00) & 0.01 (0.02, 0.00) & 0.00 (0.00, 0.00) & 0.00 (0.00, 0.00) \\ 
& BIC & 0.00 (0.00, 0.00) & 0.99 (0.98, 1.00) & 0.01 (0.02, 0.00) & 0.00 (0.00, 0.00) & 0.00 (0.00, 0.00) \\ 
& ICOMP &  0.00 (0.00, 0.00) & 1.00 (1.00, 1.00) & 0.00 (0.00, 0.00) & 0.00 (0.00, 0.00) & 0.00 (0.00, 0.00) \\
\hline\hline
\multirow{6}{4em}{Lomax Data} & KS & 0.00  & 0.00 & 0.98 & 0.00 & 0.02  \\ 
& AD  & 0.00  & 0.00 & 0.98 & 0.00 & 0.02  \\ 
& AIC & 0.00 (0.00, 0.00) & 0.00 (0.00, 0.00) & 0.99 (0.99, 1.00) & 0.00 (0.00, 0.00) & 0.01 (0.01, 0.00) \\ 
& BIC & 0.00 (0.00, 0.00) & 0.00 (0.00, 0.00) & 0.99 (0.99, 1.00) & 0.00 (0.00, 0.00) & 0.01 (0.01, 0.00) \\ 
& ICOMP & 0.00 (0.00, 0.00) & 0.00 (0.00, 0.00) & 0.95 (0.96, 1.00) & 0.00 (0.00, 0.00) & 0.05 (0.04, 0.00) \\ 
\hline\hline
\multirow{6}{4em}{Paralog Data} & KS  & 0.02  & 0.01 & 0.00 & 0.97 & 0.00  \\ 
& AD  & 0.01  & 0.01 & 0.00 & 0.98 & 0.00 \\ 
& AIC & 0.02 (0.02, 0.00) & 0.01 (0.01, 0.00) & 0.00 (0.00, 0.00) & 0.97 (0.98, 1.00) & 0.00 (0.00, 0.00) \\ 
& BIC & 0.02 (0.02, 0.00) & 0.01 (0.01, 0.00) & 0.00 (0.00, 0.00) & 0.97 (0.98, 1.00) & 0.00 (0.00, 0.00) \\ 
& ICOMP & 0.02 (0.02, 0.00) & 0.07 (0.07, 0.00) & 0.00 (0.00, 0.00) & 0.91 (0.91, 1.00) & 0.00 (0.00, 0.00) \\ 
\hline\hline
\multirow{6}{4em}{Weibull Data} & KS  & 0.00  & 0.00 & 0.12 & 0.00 & 0.88  \\ 
& AD  & 0.00  & 0.00 & 0.08 & 0.00 & 0.92 \\ 
& AIC & 0.00 (0.00, 0.00) & 0.00 (0.00, 0.00) & 0.11 (0.14, 0.04) & 0.00 (0.00, 0.00) & 0.89 (0.86, 0.96) \\ 
& BIC & 0.00 (0.00, 0.00) & 0.00 (0.00, 0.00) & 0.11 (0.14, 0.04) & 0.00 (0.00, 0.00) & 0.89 (0.86, 0.96) \\ 
& ICOMP & 0.00 (0.00, 0.00) & 0.00 (0.00, 0.00) & 0.01 (0.01, 0.00) & 0.00 (0.00, 1.00) & 0.99 (0.99, 1.00) \\ 
\hline
\end{tabular}
}
\end{table} 

The records of a best-case scenario of $F(d)=0.1$ and $F(u)=0.85$ are displayed in Tables \ref{tab:1} and \ref{tab:2}, in which 75\% of the data are used to estimate the parameters and validate the model performance. Overall, each model is distinguishable from the others. When the sample size is $n=1,000$ in Table \ref{tab:1}, the likelihood-based indicators AIC and BIC are significantly better than the distance-based indicators KS and AD on identifying correct distribution for Fisk and Lomax data, whereas the opposite performance are observed for Weibull data. In particular, when all five candidate distributions are used to fit Weibull data, KS and AD have 59\% and 50\% chances, respectively, to correctly select the underlying Weibull distribution while AIC and BIC only have around 20\% to 23\% correction rate. Meanwhile, ICOMP always specifies Lognormal model no matter what the underlying distribution, thus it is not a reliable selection criterion under this situation.  Besides, selection probabilities and posterior probabilities are generally consistent to indicate the best fit, that means, when a selection probability is large enough to specify a candidate model, the corresponding posterior probability will move to a significantly higher value than the benchmark of 20\%.\\

\begin{table}[htb!]
{\footnotesize \caption{$100$ simulations of Fisk data ($\alpha=0.46$, $\theta=500$); Lognormal data ($\mu=6.21$, $\theta=3.56$); Lomax data ($\alpha=0.32$, $\theta=63.91$); Paralogistic data ($\alpha=0.61$, $\theta=144.04$); Weibull data ($\alpha= 0.28$, $\theta=184.09$); $d=500$, $u=10,000$, $n=1,000$. $F(d)=0.50$, $F(u)=0.80$.}
\label{tab:3}
\begin{tabular}{c|c|c|c|c|c|c}
\hline
\multicolumn{2}{c|}{} &
\multicolumn{1}{|c|}{Fisk} &
\multicolumn{1}{|c|}{Logn} &
\multicolumn{1}{|c|}{Lomax} &
\multicolumn{1}{|c|}{Paralog} &
\multicolumn{1}{|c}{Weibull} \\
\hline\hline
\multirow{6}{4em}{Fisk Data} & KS  &  0.11 & 0.11 & 0.22 & 0.10 & 0.46\\ 
 & AD  &  0.13 & 0.11 & 0.26 & 0.09 & 0.41\\ 
& AIC  &  0.56 (0.20, 0.20) & 0.16 (0.22, 0.22) & 0.16 (0.17, 0.16) & 0.00 (0.19, 0.20) & 0.12 (0.22, 0.22)\\ 
& BIC   &  0.55 (0.20, 0.20) & 0.16 (0.21, 0.22) & 0.17 (0.18, 0.16) & 0.01 (0.19, 0.19) & 0.11 (0.23, 0.22)\\ 
& ICOMP &  0.00 (0.00, 0.00) & 1.00 (0.99, 1.00)& 0.00 (0.00, 0.00) & 0.00 (0.00, 0.00)& 0.00 (0.00, 0.00)\\ 
\hline\hline
\multirow{6}{4em}{Logn Data} & KS & 0.05  & 0.12 & 0.28 & 0.17 & 0.48  \\ 
& AD  & 0.06 &  0.08 & 0.31 & 0.08 & 0.47\\ 
& AIC & 0.46 (0.20, 0.21) & 0.14 (0.22, 0.22) & 0.24 (0.17, 0.14) & 0.00 (0.18, 0.20) & 0.16 (0.23, 0.22) \\ 
& BIC & 0.40 (0.20, 0.20) & 0.22 (0.22, 0.22) & 0.23 (0.18, 0.14) & 0.00 (0.18, 0.20) & 0.15 (0.00, 0.00) \\ 
& ICOMP &  0.00 (0.00, 0.00) & 1.00 (1.00, 1.00) & 0.00 (0.00, 0.00) & 0.00 (0.00, 0.00) & 0.00 (0.00, 0.00) \\
\hline\hline
\multirow{6}{4em}{Lomax Data} & KS   & 0.35 & 0.07 & 0.23 & 0.16 & 0.19 \\ 
& AD  & 0.27  & 0.05 & 0.43 & 0.11 & 0.14  \\ 
& AIC & 0.80 (0.23, 0.22) & 0.05 (0.14, 0.13) & 0.15 (0.24, 0.25) & 0.00 (0.23, 0.22) & 0.10 (0.17, 0.16) \\ 
& BIC & 0.76 (0.23, 0.22) & 0.07 (0.14, 0.13) & 0.12 (0.23, 0.24) & 0.00 (0.22, 0.22) & 0.05 (0.18, 0.18) \\ 
& ICOMP & 0.07 (0.07, 0.00) & 0.70 (0.71, 0.96) & 0.00 (0.00, 0.00) & 0.23 (0.21, 0.00) & 0.00 (0.01, 0.00) \\ 
\hline\hline
\multirow{6}{4em}{Paralog Data} & KS  & 0.11  & 0.15 & 0.28 & 0.12 & 0.34  \\ 
& AD  & 0.17  & 0.09 & 0.37 & 0.08 & 0.29 \\ 
& AIC & 0.71 (0.20, 0.20) & 0.11 (0.20, 0.20) & 0.15 (0.20, 0.20) & 0.00 (0.20, 0.20) & 0.03 (0.20, 0.20) \\ 
& BIC & 0.64 (0.20, 0.20) & 0.10 (0.19, 0.20) & 0.18 (0.20, 0.20) & 0.02 (0.20, 0.20) & 0.06 (0.20, 0.20) \\ 
& ICOMP & 0.02 (0.02, 0.00) & 0.96 (0.95, 1.00) & 0.00 (0.00, 0.00) & 0.02 (0.02, 0.00) & 0.00 (0.02, 0.00) \\ 
\hline\hline
\multirow{6}{4em}{Weibull Data} & KS  & 0.08  & 0.09 & 0.05 & 0.14 & 0.64  \\ 
& AD  & 0.04  & 0.13 & 0.07 & 0.09 & 0.67 \\ 
& AIC & 0.39 (0.19, 0.20) & 0.16 (0.24, 0.23) & 0.04 (0.10, 0.08) & 0.01 (0.16, 0.15) & 0.40 (0.31, 0.29) \\ 
& BIC & 0.35 (0.19, 0.20) & 0.16 (0.24, 0.24) & 0.006(0.10, 0.08) & 0.01 (0.16, 0.15) & 0.42 (0.31, 0.29) \\ 
& ICOMP & 0.00 (0.00, 0.00) & 1.00 (1.00, 1.00) & 0.00 (0.00, 0.00) & 0.00 (0.00, 0.00) & 0.00 (0.00, 0.00) \\ 
\hline
\end{tabular}
}
\end{table} 

As the sample size increases to $n=100,000$, all measures work very well (with at least 88\% selection accuracy) in Table \ref{tab:2}. Although there is no significant difference on selection performance, we observe that AD and ICOMP are slightly better for identifying Weibull data. Posterior probability and selection probability are almost identical within each likelihood approach comparison and the median posterior even could specify a correct parent distribution with 100\% probability.\\

\begin{table}[htb!]
{\footnotesize \caption{$100$ simulations of Fisk data ($\alpha=0.46$, $\theta=500$); Lognormal data ($\mu=6.21$, $\theta=3.56$); Lomax data ($\alpha=0.32$, $\theta=63.91$); Paralogistic data ($\alpha=0.61$, $\theta=144.04$); Weibull data ($\alpha= 0.28$, $\theta=184.09$); $d=500$, $u=10,000$, $n=100,000$. $F(d)=0.50$, $F(u)=0.80$.}
\label{tab:4}
\begin{tabular}{c|c|c|c|c|c|c}
\hline
\multicolumn{2}{c|}{} &
\multicolumn{1}{|c|}{Fisk} &
\multicolumn{1}{|c|}{Logn} &
\multicolumn{1}{|c|}{Lomax} &
\multicolumn{1}{|c|}{Paralog} &
\multicolumn{1}{|c}{Weibull} \\
\hline\hline
\multirow{6}{4em}{Fisk Data} & KS  &  0.67 & 0.10 & 0.00 & 0.23 & 0.00\\ 
 & AD  &  0.70 & 0.08 & 0.00 & 0.22 & 0.00\\ 
& AIC  &  0.75 (0.62, 0.72) & 0.09 (0.13, 0.02) & 0.00 (0.00, 0.00) & 0.16 (0.24, 0.18) & 0.00 (0.01, 0.00)\\ 
& BIC   &  0.77 (0.62, 0.72) & 0.09 (0.13, 0.02) & 0.00 (0.00, 0.00) & 0.14 (0.24, 0.18) & 0.00 (0.01, 0.00)\\ 
& ICOMP &  0.00 (0.02, 0.01) & 0.99 (0.93, 0.99)& 0.00 (0.00, 0.00) & 0.11 (0.05, 0.00)& 0.00 (0.00, 0.00)\\ 
\hline\hline
\multirow{6}{4em}{Logn Data} & KS & 0.08  & 0.78 & 0.00 & 0.00 & 0.14  \\ 
& AD  & 0.08 &  0.78 & 0.00 & 0.00 & 0.14\\ 
& AIC & 0.09 (0.12, 0.03) & 0.83 (0.67, 0.73) & 0.00 (0.00, 0.00) & 0.00 (0.001, 0.000) & 0.08 (0.21, 0.15) \\ 
& BIC & 0.09 (0.12, 0.03) & 0.83 (0.68, 0.73) & 0.00 (0.00, 0.00) & 0.00 (0.00, 0.00) & 0.08 (0.21, 0.12) \\ 
& ICOMP &  0.00 (0.00, 0.00) & 1.00 (1.00, 1.00) & 0.00 (0.00, 0.00) & 0.00 (0.00, 0.00) & 0.00 (0.00, 0.00) \\
\hline\hline
\multirow{6}{4em}{Lomax Data} & KS & 0.06  & 0.00 & 0.76 & 0.18 & 0.00  \\ 
& AD  & 0.04  & 0.00 & 0.76 & 0.20 & 0.00  \\ 
& AIC & 0.14 (0.14, 0.09) & 0.00 (0.00, 0.00) & 0.81 (0.59, 0.67) & 0.05 (0.27, 0.25) & 0.00 (0.00, 0.00) \\ 
& BIC & 0.14 (0.14, 0.09) & 0.00 (0.00, 0.00) & 0.81 (0.59, 0.67) & 0.05 (0.27, 0.25) & 0.00 (0.00, 0.00) \\ 
& ICOMP & 0.42 (0.29, 0.25) & 0.00 (0.00, 0.00) & 0.58 (0.44, 0.45) & 0.00 (0.27, 0.27) & 0.00 (0.00, 0.00) \\ 
\hline\hline
\multirow{6}{4em}{Paralog Data} & KS  & 0.19  & 0.00 & 0.07 & 0.74 & 0.00  \\ 
& AD  & 0.18  & 0.00 & 0.08 & 0.74 & 0.00 \\ 
& AIC & 0.23 (0.23, 0.18) & 0.00 (0.02, 0.00) & 0.11 (0.12, 0.02) & 0.66 (0.63, 0.71) & 0.00 (0.00, 0.00) \\ 
& BIC & 0.23 (0.24, 0.18) & 0.00 (0.02, 0.00) & 0.11 (0.12, 0.02) & 0.66 (0.63, 0.69) &0.00 (0.00, 0.00) \\ 
& ICOMP & 0.00 (0.03, 0.03) & 0.65 (0.57, 0.59) & 0.14 (0.13, 0.01) & 0.21 (0.28, 0.26) & 0.00 (0.00, 0.00) \\ 
\hline\hline
\multirow{6}{4em}{Weibull Data} & KS  & 0.00  & 0.02 & 0.00 & 0.00 & 0.98  \\ 
& AD  & 0.00  & 0.03 & 0.00 & 0.00 & 0.97 \\ 
& AIC & 0.00 (0.00, 0.00) & 0.03 (0.06, 0.00) & 0.00 (0.00, 0.00) & 0.00 (0.00, 0.00) & 0.97 (0.94, 1.00) \\ 
& BIC & 0.00 (0.00, 0.00) & 0.03 (0.06, 0.00) & 0.00 (0.00, 0.00) & 0.00 (0.00, 0.00) & 0.97 (0.94, 1.00) \\ 
& ICOMP & 0.00 (0.00, 0.00) & 0.90 (0.86, 0.99) & 0.00 (0.00, 0.00) & 0.00 (0.00, 0.00) & 0.10 (0.14, 0.01) \\ 
\hline
\end{tabular}
}
\end{table} 

In a worst-case scenario of $F(d)=0.5$ and $F(u)=0.8$ in Table \ref{tab:3}, we only use 30\% of middle values to estimate the parameters and identify the parent distributions. As a result, the true underlying model is generally not distinguishable from other candidates when $n=1,000$. It is clear that AIC and BIC always tend to pick up Fisk model as the best fit no matter what the parent model is except Weibull as the parent distribution. In particular, if the data {are} generated from Paralogistic distribution, there is no chance for AIC and BIC to identify it since the probabilities that Paralogistic model is chosen as the best candidate are 0\% and 2\%, respectively.
Meanwhile, KS and AD prefer to select Weibull, but this tendency overall is not as strong as the likelihood based measures AIC and BIC to select Fisk.
These outcomes give rise to a scenario of model uncertainty, wherein posterior probabilities appear to equally fluctuate around 20\%, leading to no signal for model preference. As the sample size is increased to $100,000$ in Table \ref{tab:4}, all measures except ICOMP work reasonably well for separating models. The correct selection rate, especially for Paralogistic data, has moved up from 0\% to 66\% in likelihood-based measures, although there is still an 8\% gap to the distance-based indicators. 

\section{Real Data illustration}
Next, a real data analysis is conducted to illustrate the procedure of parametric model selection when the loss deductible and policy limit are under consideration.  The target data {are} the Local Government Property Insurance Fund (LGPIF), an insurance pool administered by the Wisconsin Office of the Insurance Commissioner, which has been studied by \cite{Frees2018}. This insurance fund covers local government properties that include counties, cities, towns, villages, school districts, and library boards. Our goal is to investigate what effect initial assumptions have on the MLE estimates with various  typical loss models and how to pick up an appropriate candidate to predict the claims for potential risk management.\\

\begin{figure}[htb!]
    \centering
    \includegraphics[scale=0.9]{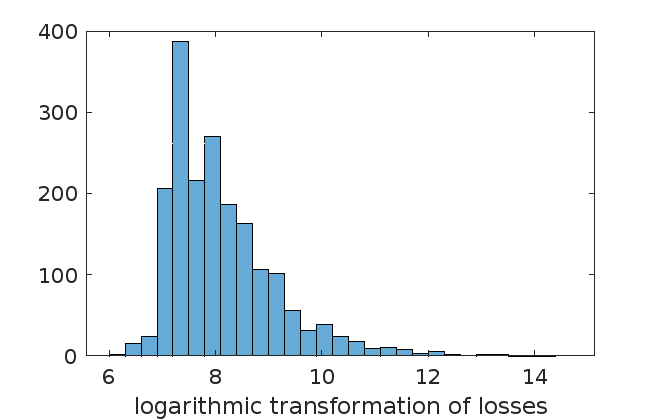}
    \caption{Loss data of LGPIF when the deductible is 500}
    \label{fig:loss}
\end{figure}

With a 500 deductible ($d=500$), 1890 LGPIF loss payments were recorded in the portfolio and the histogram of losses was displayed in Figure \ref{fig:loss}.
It is clear that the loss values resemble a heavy-tailed distribution even after the logarithmic transformation. The above-mentioned Fisk, Frechet, Lognormal, Lomax, Paralogistic, and Weibull could be good candidates to investigate the uncertainty of claim payment for this insurance fund.\\

This fund covers seven different types of property. Due to the various ranges of loss magnitude, we set a flexible policy limit for each individual type, stopping at the 80th percentile of the loss (that is $F(u)=0.8$), respectively. The estimated parameters and the model selection indicators are recorded in Table 6.1. All values of AD, AIC and BIC are the lowest for Frechet model and the differences in measure indicators are tremendously significant. This indicates a signal to prepare for risk management, that is, LGPIF may have an extremely heavy tail to their potential claim payment.\\

\begin{figure}[hbt!]
\begin{subfigure}{.5\textwidth}
\hspace*{0.45 in}
\includegraphics[height=5.6cm,width=7.8cm]{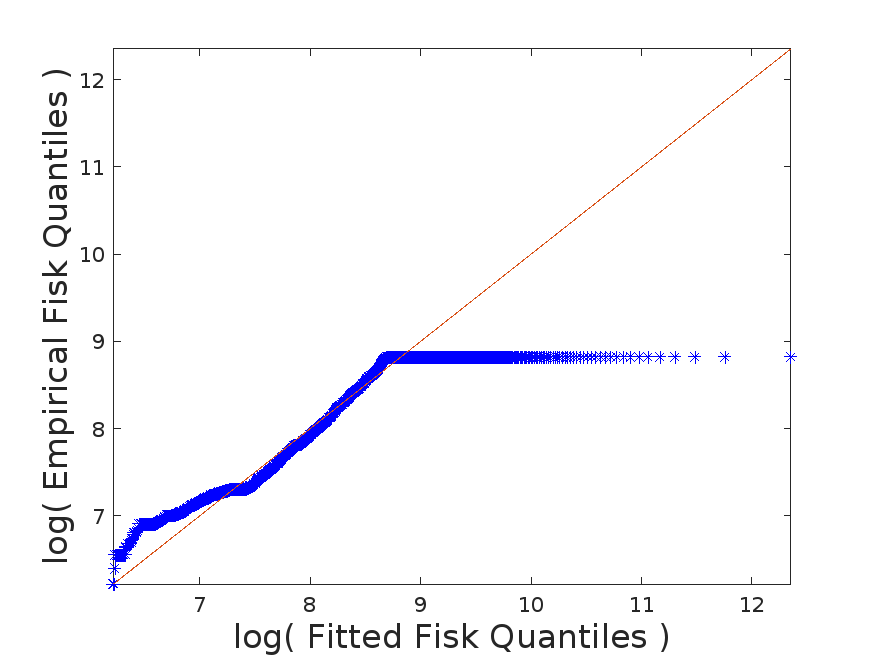}
\end{subfigure}
\quad
\begin{subfigure}{.5\textwidth}
\hspace*{-0.25 in}
\includegraphics[height=5.6cm,width=7.75cm]{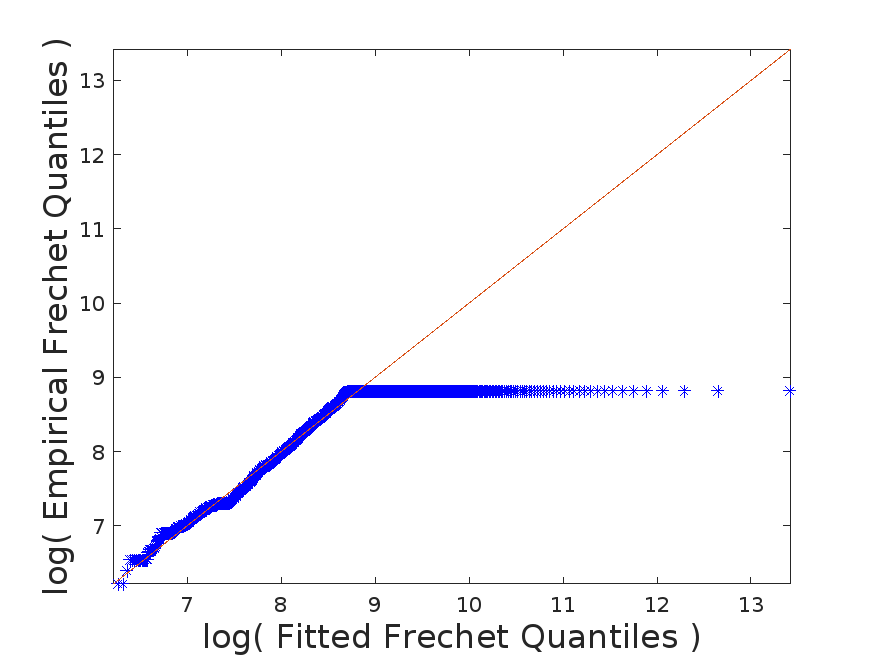}
\end{subfigure}
\qquad
\begin{subfigure}{.5\textwidth}
\hspace*{0.5 in}
\includegraphics[height=5.5cm,width=7.7cm]{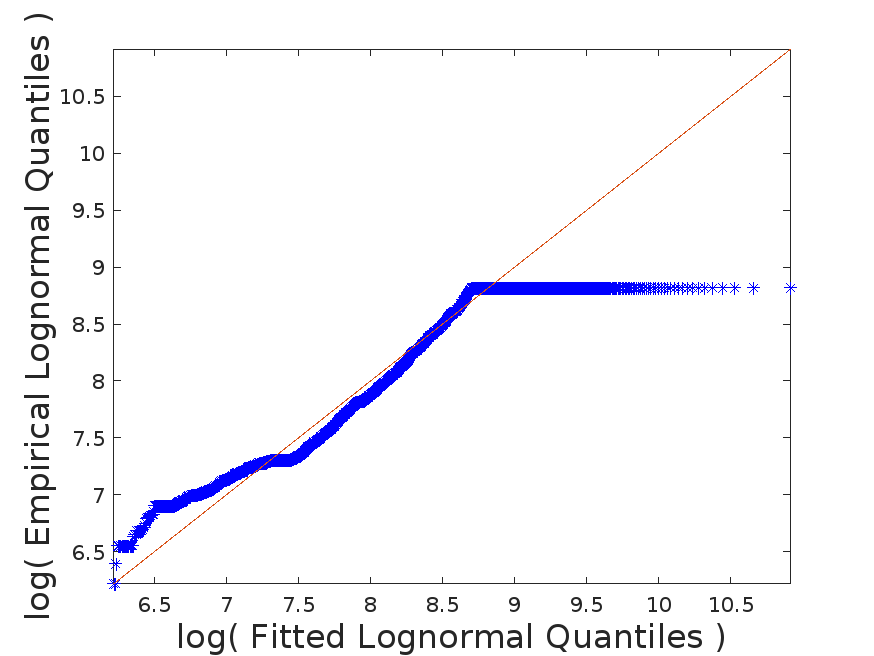}
\end{subfigure}
\begin{subfigure}{.5\textwidth}
\hspace*{-0.045 in}
\includegraphics[height=5.5cm,width=7.7cm]{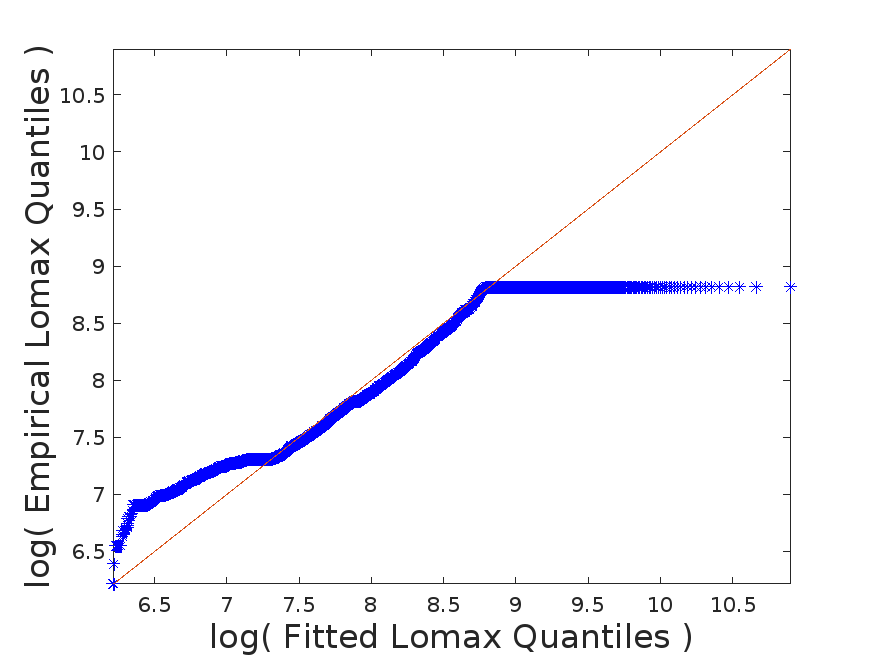}
\end{subfigure}
\qquad
\begin{subfigure}{.5\textwidth}
\hspace*{0.5 in}
\includegraphics[height=5.5cm,width=7.7cm]{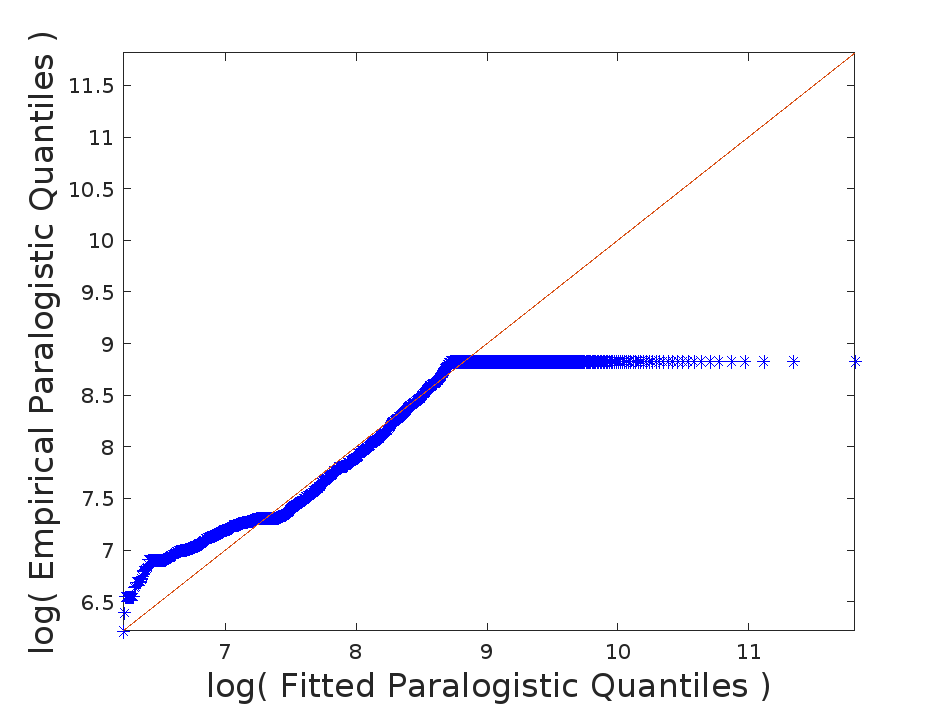}
\end{subfigure}
\begin{subfigure}{.5\textwidth}
\hspace*{-0.08 in}
\includegraphics[height=5.5cm,width=7.7cm]{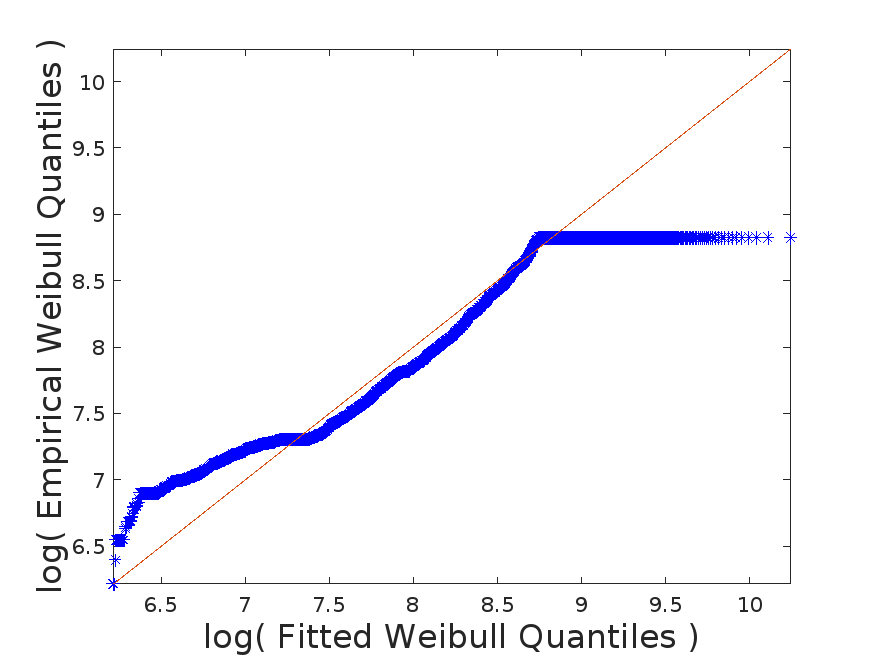}
\end{subfigure}
\caption{Q-Q plot of the fitted models to the real data example.}
\label{fig:QQ}
\end{figure}

\begin{table}[htp]
\begin{center}
{\footnotesize
\caption{Real data illustration with $d=500$, and empirical $F(u)=0.8$.}
\label{tab:real}
\begin{tabular}{c|c|c|c|c|c|c}
\hline
& Fisk &  Frechet &
Logn &
Lomax  &
Paralog &
Weibull \\
\hline
$\hat{\theta}$ & 2744.4 & 2125.4 & 7.9832 & 21,999 &  3912.4& 3731.9\\
$\hat{\alpha}$ & 1.8661 & 1.4296 & 0.8453 & 6.7464 & 1.5303 & 1.051\\
KS &  0.0741 &   0.0678 &    0.0665 &  0.1161 &  0.0865 &  0.0999\\
AD &   15.4702& 6.6583 &  17.6513 &  30.6288 & 19.2924& 27.5664
\\
AIC & 27,669 & 27,472
&   27,664  &  27,821 &  27,720 & 27,828\\ 
BIC & 27,680 &  27,483 &   27,678
& 27,833 & 27,731 & 27,839\\ 
ICOMP & 27,678 &  NA
  & 27,663 &  27,838 &  27,732 & 27839\\ 
 $\widehat{F(d)}$, $\widehat{F(u)}$ & 0.04, 0.8440 & 0.0004, 0.8267 & 0.0182, 0.8395 & 0.1407,  0.8368 &  0.0623, 0.8407 & 0.1139,0.8465\\ 
\hline
\end{tabular}
}
\end{center}
{\footnotesize
Note: $\hat{\theta}$ and $\hat{\alpha}$ are estimates based on the formulas in Section 2. For Lognormal distribution, $\hat{\theta}$ represents $\hat{\mu}$ and $\hat{\alpha}$ represents $\hat{\sigma}.$
}
\end{table} 
We also present plots of the fitted-versus-observed quantiles 
for the Fisk, Frechet, Lognormal, Lomax, Paralogistic, and Weibull distributions in Figure \ref{fig:QQ}. Clearly, Frechet-estimated quantiles fall almost perfectly on the $45^{\circ}$ 
line against the empirical quantiles. On the other hand, Fisk, Lognormal, Lomax, Paralogistic, and Weibull QQ plots do not look as good, and having bottom observations above the $45^{\circ}$ line indicates that 
these models underestimate the right tail of the data.\\

\section{Conclusion {and Discussion}}

We have presented our research on model uncertainty and selection in the modeling of insurance loss data using left-truncated and right-censored (LTRC) distributions. We have considered six candidate distributions: Fisk, Frechet, Lognormal, Lomax, Paralogistic, and Weibull. Maximum Likelihood Estimation (MLE) based approaches are used to estimate the parameters of the distributions under consideration.\\

A variety of model validation methods, including Quantile-Quantile (QQ) plots, Kolmogorov-Smirnov (KS) and Anderson-Darling (AD) statistics are employed. These tools are critical for assessing how well each distribution fits the data. Model selection adopts information criteria such as Akaike Information Criterion (AIC), Bayesian Information Criterion (BIC), and Information Complexity (ICOMP). These criteria help us choose the most suitable model while penalizing the complexity of the model.\\

A simulation study has been conducted to evaluate the performance of each model validation method and model selection criterion under different scenarios of left truncation and right censoring. This is achieved through calculating both selection probability and posterior probability for each candidate model. Uncertainty has been observed in the patterns regarding which model validation method or model selection criterion is outperforming others, depending on a lot of factors, such as sample size, the underlying true distribution, left truncation proportion, and right censoring proportion, etc. \\

In the simulation study, we have noticed that multiple models may be difficult to distinguish, and various tests and/or criteria may lead to the selection of different models being the best. It seems that distance-based criteria tend to select Weibull model and likelihood-based criteria prefer to select Fisk model among the considered candidate models. Then model averaging can be considered at this time. Model averaging can provide more robust estimates, especially when there is uncertainty in identifying the true underlying distribution. \\

Analyzing actual local government property loss data in Wisconsin serves as a practical application of our modeling methodology. Based on the outputs of the model selection indicators and the QQ plots, we identify Frechet distribution as the best fit to LGPIF, which is a significant finding, especially given its relevance for modeling extreme events. \\

In summary, our study combines theoretical and empirical approaches to analyze LTRC insurance loss data. The use of various statistical tools and criteria helps ensure the reliability of our modeling results, which exhibit a certain degree of uncertainty in the model selection procedure illustrated through both simulation and case studies. Through our research in this paper, we would like to give the following recommendations in regard to uncertainty in the model selection procedure, using our selection probability framework. If KS and AD can provide a strong evidence to select a best candidate model, then it is very likely that AIC and BIC will come up with the same decision. When KS and AD can only provide a fairly weak evidence for the selection of a best model, then we may consider further use of AIC and BIC to reinforce the decision if a common best model has been selected by AIC/BIC as KS/AD. Otherwise, when the use of AIC and BIC leads to the selection of a different best model from the use of KS and AD, we may consider model averaging of two or more competing best models. \\

{Our comprehensive study of modeling LTRC insurance loss data addresses various aspects of model uncertainty and selection, but there are also areas where some limitations and potential avenues for future research can be discussed. \\

There are some limitations to the assumptions about candidate distributions. Our study considers six candidate distributions for modeling insurance loss data. However, it is worth noting that these distributions may not capture all possible underlying distributions. There are other limitations to the sensitivity analysis. We have already mentioned that the performance of model validation methods and selection criteria will vary with factors such as sample size, left truncation and right censoring ratios. It would be useful to further quantify specific sensitivity levels. This information can help practitioners better understand the applicability of our methodology in various real-world scenarios. \\

In addition to exploring other distributions and conducting more quantitative sensitivity analysis, future research avenues could include, but are not limited to, robust model averaging, out-of-sample testing, incorporation of expert opinions, quantification of model uncertainty, and applications in other areas. First, we mention that model averaging can provide more robust estimates when there is uncertainty in identifying the true underlying distribution. Future research could delve deeper into methods and techniques for robust model averaging, exploring different weighting schemes or Bayesian model averaging. Second, while we have conducted simulation studies and applied our methods to real data, future research could focus on out-of-sample testing to evaluate how well the selected models generalize to new, unseen data. Third, incorporating expert opinions or domain knowledge into the model selection process can improve the relevance and accuracy of the selected distributions in practical applications. Fourth, it seems interesting to develop techniques that more explicitly quantify and visualize model uncertainty which can help decision-makers understand the range of potential outcomes based on different candidate distributions. Finally, we can explore how our modeling approach can be applied to areas beyond insurance loss data, such as finance, healthcare, or environmental modeling. \\

Incorporating these limitations and future research considerations will help enrich our work and provide a more complete understanding of the challenges and opportunities of modeling insurance loss data using LTRC distributions. It will also enhance the practical applicability of our findings. }

\section*{Acknowledgements}

The authors are very appreciative of valuable insights and useful comments provided by {two anonymous referees and} Professor Vytaras Brazauskas, which helped to improve the paper significantly.

\baselineskip 5mm

\bibliography{ArXiv_Main}
\thispagestyle{plain}
\pagestyle{plain} 
\thispagestyle{plain}

\end{document}